\documentstyle[draft]{l-aa}

\thesaurus{}
\newcommand{\etal}{{\it et al.}}

\newcommand{\ltappeq}{\mathrel{\hbox{\rlap{\hbox{\lower4pt\hbox{$\sim$}}}\hbox{$<$}}}}
\newcommand{\gtappeq}{\raisebox{-0.6ex}{$\,\stackrel
{\raisebox{-.2ex}{$\textstyle >$}}{\sim}\,$}}
\newcommand{\mstar}{\mbox{$M_{\ast}\;$}}
\newcommand{\rstar}{\mbox{$R_{\ast}\;$}}
\newcommand{\msunyr}{\mbox{M$_{\odot}$\thinspace yr$^{-1}\;$}}
\newcommand{\msun}{\mbox{$M_{\odot}\;$}}
\newcommand{\rsun}{\mbox{$R_{\odot}\;$}}
\newcommand{\kms}{\mbox{\ km\ s$^{-1}$}}
\newcommand{\aanda}{{\sl A\&A}}
\newcommand{\apj}{{\sl ApJ}}
\newcommand{\mnras}{{\sl MNRAS}}
\newcommand{\pasj}{\sl Publ. Ast. Soc. Japan}
\newcommand{\araa}[1]{{\sl Ann. Rev. Astr. Astrophys.} {\bf #1}}
\newcommand{\pasp}[1]{{\sl PASP} {\bf #1}}
\newcommand{\apjs}{{\sl ApJS}}

\title{On outflowing viscous disc models for Be stars}
\author{John M. Porter}
\institute{Astrophysics Research Institute,
Liverpool John Moores University,
Twelve Quays House, Egerton Wharf, \\
Birkenhead L41 1LD, UK
\thanks{email : {\tt jmp@astro.livjm.ac.uk}}
}

\date{Received 19th March, 1999; accepted 22nd June, 1999}

\begin{document}
\maketitle

\begin{abstract}
It is assumed that Be star discs are driven by viscosity. 
Emission from disc models is calculated and is confronted with
continuum observations. It is found that the outflowing viscous
disc models can reproduce the observed IR continuum emission. 
However, to exist as {\em outflowing} discs, either 
the discs are significantly
acted upon by the stellar radiation field and/or there is significant
cooling with radius in the disc.
The energy generated via viscous dissipation is calculated and shown
to play only a minor r\^{o}le in the energy balance of the disc. 

A scenario whereby a B star may change into a Be star (and vice versa)
by generating (reaccreting) the disc is suggested.


\end{abstract}

\section{Introduction}

Be stars are fast rotating stars (Slettebak 1982) with
rotation rates up to break-up, and a modal value of $\sim 70$\% of
break-up rotation (Porter 1996).
The circumstellar matter around Be stars is thought to consist of two
distinct regions:
a diffuse polar stellar wind and a dense equatorial ``disc''
(Dachs 1987, Slettebak 1988).
The nature of the polar wind seems to be well understood in the
context of radiation-driven wind theory (Castor, Abbott \& Klein 1975,
Kudritzki \etal\ 1989). However, the mechanism generating the
equatorial disc is more elusive: several theories have been proposed
including wind compression (Bjorkman \& Cassinelli 1993), outflowing viscous
discs (Lee \etal\ 1991), and wind bistability (Lamers \&
Pauldrach 1991). 
Recently it has been found that stars with high
rotation will generate a metallicity enhancement in their equatorial
planes (Porter 1999), which through the metallicity dependence of the
radiation driving parameters may change the nature of the wind there.

Wind compression has already been shown to be insufficient on
theoretical (Owocki, Cranmer \& Gayley 1996) and observational (Porter 1997)
grounds. The major problem with wind bistability is that the disc
appears to be rotating in a Keplerian fashion (e.g. Dachs \etal\
1986, Hanuschik 1989, 1996), whereas any wind causing the equatorial
enhancement is likely to conserve angular momentum, and should not be
observed to be rotating {\em faster} than the star -- Hanuschik (1996)
has shown that the half-line widths of emission lines created in the
disc are indeed larger than $v\sin i$ of the star.

The theory which seems to be most applicable to Be
star discs is that of viscously driven outflow (Lee \etal\ 1991). Here
angular momentum is added to the inner edge of the disc increasing its
angular velocity to slightly super-Keplerian. The disc interacts with itself via
viscosity and angular momentum is transported outwards. The angular
momentum source has been suggested by Osaki (1986) to be due to non-radial
pulsations dissipating in the atmosphere of the star. The outflowing viscous
disc model has several circumstantial corroborations -- the
variation in the line asymmetry (the V/R variation) can be naturally
generated in Keplerian discs (Papaloizou \etal\ 1992, Okazaki 1991). 
Excess IR emission generated by the disc is seen to disappear over a
timescale of months (Hanuschik \etal\ 1993) which is similar to the viscous
timescale in these discs -- this is perhaps indicative of phases of outflow
and inflow within the disc.

Although the outflowing viscous disc model appears to be the most
likely, it has never been subjected to observational tests -- can
these discs generate the IR excess? What is the source of
viscosity? How and why do they dissipate and regenerate?
In this investigation, Be star discs are assumed to be outflowing 
viscous discs and an attempt is made to answer these questions.

In \S2 aspects of the outflowing viscous disc model are considered along
with their application to Be star discs.
In \S3 the energy balance within the disc is studied, and the theory
is applied to a test case in \S4. Discussions and conclusions are
presented in \S5 and \S6.

\section{Can outflowing viscous discs exist?}

\subsection{Viscous discs}
The gas dynamics of viscous discs has been analysed and developed by
several authors (Shakura \& Sunyaev 1973, Pringle 1981, Frank, King \&
Raine 1992). 
Here, some comments are made regarding the structure of such discs
(following arguments of the references above).
First, a disc interacting with itself via viscous stresses transports
angular momentum outward. In order for viscous stresses to exist in a
disc, there must be (i) a non-zero viscosity and (ii) some
differential rotation in the disc. Conservation of angular momentum
in a cylindrical co-ordinate system ($R, \phi, z$),
yields
\begin{equation}
R\frac{\partial}{\partial t} \left(\Sigma R^2 \Omega\right) +
\frac{\partial}{\partial R} \left(R \Sigma v_R R^2\Omega\right) = 
\frac{1}{2\pi}\frac{\partial G}{\partial R}
\end{equation}
where $\Sigma$ is the surface density of the disc, $\Omega$ is the
angular velocity, and $v_R$ is the radial ``drift'' velocity.
The viscous torque is $G = 2\pi R\nu \Sigma R^2 (\partial
\Omega/\partial R)$ where $\nu$ is the viscosity.
If a ring of matter is
placed in a Keplerian orbit and allowed to interact with itself
through viscous stresses then part of it will move to smaller radii
having $v_R <0$ and parts of it will move outward $v_R >0$
(e.g. Pringle's 1981, fig.1). This is solely due to the redistribution
of angular momentum in the gas. The outflowing (inflowing) parts have
increased (decreased) angular momentum with respect to the initial
state. Time steady
($\partial/\partial t = 0$) solutions which show no drift velocities
$v_R = 0$ occur when $G$ is a constant with radius, i.e. $\partial
\Omega/\partial R \propto (\nu \Sigma R^3)^{-1}$, although these
solutions have limited physical application in discs around stars.
These static discs still transport angular momentum outwards, even
though the gas in them is not moving radially.

For an accretion disc, matter is added at the outer part of the disc. 
The viscous stresses lower the angular momentum
of a ring of gas which in response moves inward and increases its
angular velocity until it it regains rotational support. This
mechanism creates the accretion flow:
the gas moves inward $v_R <0$ throughout the disc.
The disc's velocity field is typically rotationally dominated with
rotational velocities very close to Keplerian, and a subsonic radial velocity.

Eq.1 also permits solutions with positive radial drift
velocity $v_R$ everywhere i.e. outflowing discs. This sort of disc
requires an angular momentum source at the inner boundary 
-- the disc-star interface -- as well as a gas source taken to be the
atmosphere. The disc
recieves angular momentum from the star
which will itself spin down. It is exactly this sort of model which has been
suggested by Lee \etal\ (1991) to apply to the
discs of Be star and are investigated below.
Porter (1998a) has already condisered the angular momentum evolution
of Be stars for this sort of disc, and finds that the observational
result that there is little or no angular momentum evolution of Be
stars (Zorec \& Briot 1997, Steele 1999) may be explained if the
viscous stresses at 
the inner edge of the disc are small or that the disc is present
intermittently in the lifetime of the star. A small couple applied
from the star at the inner edge leads to slow outflow velocities of
$v_R \ltappeq 0.01$\kms at the inner edge of the disc. Note that an
outflowing disc cannot exist if there is no couple from the star at
the inner edge. 

Aside from Lee \etal's (1991) initial suggestion and modelling of Be
star discs, outflowing viscous discs have been discussed by Pringle
(1991), Narita, Kiguchi \& Hayashi (1994) and Okazaki (1997).
Narita \etal's numerical
analysis showed that if the central object rotates fast enough, then
the disc does indeed become an outflowing one.
Let us assume then that Be star discs {\em are} outflowing viscous
discs and subject them to some observational tests.

\subsection{Angular momentum transport in a viscous disc}

The existence of an outflowing disc is dependent on angular momentum
being added at the inner edge of the disc whereas the dynamics of the
disc is determined by the viscous stresses: if only this initial couple
is present then a ring of gas will move away from the star until it finds
its Keplerian orbit, at which point it will stop. The evolution of the
disc away from the inner boundary is governed by the viscosity.

The rotational velocity of a outflowing viscous disc is
$v_\phi \approx \sqrt{G\mstar/ R}$, where \mstar is the mass
of the star, and $G$ is the gravitational constant.
Assuming that the disc is steady (so $\partial /\partial t = 0$), the
equation for transport of angular momentum becomes
\begin{equation}
\Sigma v_R R^{1/2} = -3 \frac{\partial}{\partial R}\left( R^{1/2} \nu
\Sigma \right)
\label{angmom}
\end{equation}
where $\Sigma$ is the surface density of the disc, $v_R$ is the radial
velocity, and $\nu$ is the viscosity.

The surface density is $\Sigma = \rho(R,0) H$, where $\rho(R,0)$ is the
density in the equatorial plane at radius $R$, and 
$H~=~R~c_s~/~v_\phi$ is the density scale height ($c_s$ is the sound
speed). 
The disc is isothermal in the $z$ direction 
and assuming that the density in the equatorial
plane $\rho(R,0)$ is a power law, the density field is
\begin{equation}
\rho(R,z) = \rho_0 \left(\frac{R}{\rstar}\right)^{-n} 
\exp{\left(-\frac{z^2}{2H^2}\right)}
\end{equation}
where $\rho_0$ is the density at the inner boundary.
Note that the numerical simulations of outflowing discs by Narita
\etal\ (1994) find that steady state outflowing discs do indeed have
power law surface densities from $R \gtappeq 2\rstar$, justifying the
power-law assumption in eq.3.

Following convention, an alpha prescription is used for the viscosity
$\nu = \alpha c_s H$ (Shakura \& Sunyaev, 1973).
Finally, to ensure the disc is as general as possible, the temperature
of the disc is assumed to follow a power law: $T_d = 0.8T_{\rm eff}
(R/\rstar)^{-m}$, where $T_{\rm eff}$ is the effective temperature of
the star.

\subsection{Inflow or outflow?}
Let us assume that we have a disc which is viscously interacting with
itself. Is it possible to tell whether the gas flow is inward, outward or
zero? In asking this question, the nature of the addition of
torque to the inner boundary of the disc for outflowing discs, or that
of the angular momentum removal for inflowing discs has been
ignored. An attempt is made to ascertain whether outflowing viscous
discs are credible Be star disc candidates.

For $v_R > 0$
then the radial exponent of the term in brackets in eq.\ref{angmom}
must be less than zero, ensuring the differential is negative and
hence the right hand side is positive.  
Inserting eq.3 and the $\alpha$ viscosity into eq.\ref{angmom} and
collecting powers of radius $R$
together, then
the right hand side is proportional to $-\partial/\partial R(R^{3.5 -
n - 1.5m})$.  Therefore the disc is an outflowing disc if $2n + 3m >
7$, assuming that $\alpha$ is a constant (see later).
It is worthy of note that when outflowing viscous discs have been
modelled, this limit on $n$ and $m$ has always been obeyed (e.g. see
Narita, Kiguchi \& Hayashi 1994 and Okazaki 1997).


\subsection{Confrontation with observations}

There does appear to be a problem with the outflowing viscous disc paradigm
straight away when confronting it with observations. 
C\'{o}te \& Waters (1987) found that
the IR emission (originating
in the inner 10s of \rstar of the disc) 
is well fit with an isothermal disc ($m = 0$) having a
density power law index of $2 < n < 3.5$.
Values of $n$ are larger when fitting the emission 
in the radio regime, corresponding to larger radii of the disc
(Dougherty \etal, 1991).
According to the above work, 
the inner parts of the discs {\em cannot} be outflowing viscous discs,
and must actually be accretion discs!

If the outflowing viscous disc model is applicable to Be star discs
then this discrepancy {\em must} be examined.
First, consider the emission fitting procedure: typically
both a constant opening angle for the disc and an isothermal disc are used.
The constant opening angle leads to $H \propto R$ as opposed to
$H\propto R^{3/2}$ for viscous discs above. 
Inserting this slight modification into the angular momentum conservation
expression, leads to the new limit $2n + 3m > 6$ for outflow. 

However, comparing this with the derived values of $n$ does
not really change the conclusion that almost
all discs should be accreting!
We must be wary of taking the
fits at face value -- usually the disc was assumed {\it a priori} to be
isothermal, and the parameter $m$ was not allowed to vary. If a
non-isothermal disc is used then the fits to the IR excess may change.

What can be gleaned from this apparent paradox that all Be star discs
should be inflowing, given that there is a strong belief that they are
in fact outflowing? 
It is unlikely that the fitting
procedure produces values of $n$ consistently too small in almost
every cases (see \S4). 
The conclusion, therefore, is that the outflowing viscous disc can not
fit the observations, and cannot be applicable to Be stars.

This is too hasty: aspects of the problem have yet to be considered.
The first is the radiation field from the star: 
the total electron-scattering optical depth in the equatorial plane is
\begin{eqnarray}
\tau & = & \int^{\infty}_{\rstar} \sigma_e \rho dR =
\frac{\sigma_e \rho_0 \rstar}{n} \\ 
     & = & \frac{0.24}{n}
\left(\frac{\rho_0}{10^{-11}{\rm g\ cm}^{-3}} \right)
\left(\frac{\rstar}{\rsun}\right) \nonumber
\label{tau}
\end{eqnarray}
where $\sigma_e = 0.35$cm$^2$g$^{-1}$ is the electron scattering
cross-section. Inserting typical values of $n=2.5$, $\rho_0 =
10^{-11}$g cm$^{-3}$ and $\rstar=5\rsun$
yields an optical depth of $\tau = 0.5$. Therefore the disc is
optically thin to electron scattering, and hence the disc may be acted
upon by optically thin lines, as suggested by Chen \& Marlborough
(1994) (note that the optical depth argument derived in \S4.1 of Chen \&
Marlborough where they rule out radially subsonic discs is misleading
due to their choice of wind parameters: viscous discs may have
mass-loss rates of 10$^{-11}$\msunyr with small radial velocities and
still be dense enough to produce the IR excess emission e.g. Okazaki 1997).

If regions of the disc are partially supported and driven outwards by
radiation, then they will deviate from Keplerian rotation. This
is because an insufficient amount of angular momentum is given to a
ring of material as it moves outward. Hence the rotation velocity of
the disc will lie between Keplerian ($v_\phi\propto r^{-1/2}$) and angular
momentum conserving ($v_\phi\propto r^{-1}$). This difference may be very
difficult to discriminate observationally, and so radiation-driven
discs cannot be ruled out. 
However, it should also be noted that the inclusion of optically thin
radiation driving does not necessarily produce non-Keplerian discs
(e.g. see Okazaki 1997).

It is noted that the viscosity parameter $\alpha$ may be dependent on
radius. 
The parameter $\alpha$ represents the difference between the {\em actual}
value of (size$\times$velocity) of the turbulent eddies and the
characteristic product of scale and speed in the disc (disc height$\times$sound
speed). 
In fact if the turbulence becomes supersonic then it is possible that
$\alpha > 1$. 
In the situation currently considered, a dense rotationally-dominated
disc is adjacent to the fast wind which is dominated by the radial
component of velocity. In the interaction region there are large
shearing velocities and therefore the interaction region will become
unstable to Kelvin-Helmholtz instabilities. 

Consequently it might be expected that the typical turbulent velocity
will be a function of the Mach number of the fast wind. 
In fact, if the eddy size is the typical scale height of the disc,
then the maximum value that is possible if this is the case
will be the fast wind Mach number -- the wind typically reaches
velocities of $>$1000\kms\ after a few \rstar, and with the typical
temperature of the disc of 10\kms, yield Mach numbers of ${\cal M}
\sim 100$. Therefore, it is conceviable that may become larger than
unity at quite a small radius, although $\alpha < {\cal M} \approx 100$. 

As the fast wind's velocity increases with distance from the star,
then $\alpha$ might be expected to
increase with radius. This aspect of the viscosity is difficult
to examine and a full study is left to a future paper.
It is expected that if this Kelvin-Helmholtz process is significant in
determining $\alpha$, then it should increase with radius. This,
however, makes the gas in the observed discs more likely to be inflowing
rather than outflowing. 

It is also noted that the disc may be outflowing if there is a
sufficient radial temperature gradient in the disc (corresponding to
large $m$). This requires the disc to cool significantly as it drifts
away from the star. However, recent work by Millar \& Marlborough (1998) casts
doubt on a such a large temperature gradient being present.

\subsection{Accretion discs?}
The above paragraphs attempt to show that
outflowing viscous discs are acceptable models for Be star discs.
However, to make that discussion the primary one for the rest of this
paper, we must answer the question ``why are Be star discs not normal
accretion discs?'' 

There are two prerequisites for them to be accretion discs: there must
be a supply of gas to make the disc, and the gas must have larger
specific angular momentum than Keplerian at the disc surface. 
The first of these two aspects can be fulfilled at least for some
stars with weak radiatively driven winds: Porter \& Skouza (1999)
have examined models of radiatively driven winds where the gas and
radiation field decouple due to ion stripping (e.g. Springmann \&
Pauldrach 1992), before the flow becomes unbound from the star. The
gas then stalls in the star's gravitational potential and
reaccretes.
This may provide the material to make an accretion disc if the shell
then was concentrated in the equatorial plane, but it will
not have enough specific angular momentum to {\em form} an accretion
disc. If the gas conserves angular momentum (almost certainly
true for the outflowing phase), then when the gas reaccretes, it will
have the same angular velocity as it left the star,
i.e. sub-Keplerian. Consequently, unless there is some way of adding
angular momentum to the wind as it stalls, this does not seem a
feasible way to make accretion discs.
A second possible reservoir of gas is a companion star, which is losing
mass to the Be star -- this does not have the angular
momentum problem, although it does mean that all Be stars are in
binary systems.

On the balance of these points, it seems unlikely that Be stars discs
are {\em accretion} discs
because (i) not all Be stars are in interacting binary systems with
the secondary star as the mass donor, and (ii) an accretion disc will
not form when a star is reaccreting its own wind.

\section{Energy balance within the disc}

\subsection{Liberation of energy via viscous dissipation}

The viscous stresses enables the disc to flow outwards and in doing so 
liberates energy. What happens to this energy?
The timescale over which the viscous disc can evolve significantly
$\tau_v$ is defined: 
\begin{eqnarray}
\tau_v & \sim & \frac{R^2}{\nu} = \frac{(G\mstar R)^{1/2}}{\alpha
c_s^2} \\
       &  \approx   & 21 \left(\frac{\mstar\rstar}{\msun\rsun}\right)^{1/2} 
\frac{1}{\alpha T_4} \ \ {\rm days} \nonumber
\end{eqnarray}
where $T_4$ is the disc
temperature in 10$^4$K. Inserting typical parameters leads to $\tau_v
\sim 10^2$s of days. Given that Be star discs are observed to be present
for many years (a pre-requisite for them to be able to develop and evolve V/R
variations in the lines), then a typical disc is stable over a viscous
timescale. Therefore any energy liberated via viscous stresses {\em must}
emerge as radiation.

The luminosity emitted is
\begin{eqnarray}
\frac{dL}{dR} & = & \frac{9\pi}{2} G \mstar \nu \Sigma R^{-2}\nonumber \\
              & = & \frac{9\pi}{2} \rho_0 c_s^3 \rstar \alpha 
\left( \frac{R}{\rstar}\right)^{1-n}
\label{dldr}
\end{eqnarray}
(Frank, King \& Raine, 1992, p73), where the expressions for $H$,
$\Sigma$ and $\nu$ have 
been inserted to obtain the second equality.
This luminosity source is termed the ``viscous luminosity'' as it is
the energy emitted caused solely by the viscous processes leading to
outflow. 
This expression only gives the luminosity as a function of radius, and
contains no spectral information. As the disc is optically thin in the
$z$ direction, then
the standard approach to the disc emission (e.g. Section 5.5 of Frank,
King \& Raine 1992)
are inapplicable, and an alternative discussion is needed.

\subsection{IR free-free and free-bound emission}

The continuum emission of Be star discs is limited to the
IR--radio spectral regions (e.g. see fig.11 of Poeckert \& Marlborough 1978)
and was identified as free-free and free-bound emission from an ionized
plasma by Gehrz \etal\ (1974). The most successful empirical model
used to calculate the emission is due to Waters (1986), and
represents the disc using a density power law, and an opening
angle.
Three parameters determine the emission: the opening angle, the
density at the star-disc boundary, and the density power-law exponent.

The above expressions for a outflowing viscous disc changes the
calculation of disc emission and so some of Waters' expressions
are now restated.
The optical depth $\tau_\nu$ of the disc may be calculated at a
frequency $\nu$. The expression is simpler than Waters' expression as
the integral through the disc (equivalent to his $C(\theta,n)$) may be
completed analytically. Using Waters' notation 
\begin{equation}
\left.
\begin{array}{c}
\tau_\nu(R) = X_\lambda\ X_{\ast d}\ R^{-2n+3/2 - m/2} 
\sqrt{\frac{\displaystyle \pi c_{s,0}^2 \rstar}{\displaystyle G \mstar}} \\
  \\
X_\lambda  = \lambda^2 
\frac{\left(\displaystyle {1 - {\rm e}^{-h\nu / kT_d }}
\right)}{\left(\displaystyle {\frac{h\nu}{kT_d}}\right)} \left( g_{ff} + g_{fb}
\right) \\
 \\
X_{\ast d} = 4.923\times 10^{35} {\displaystyle {\frac{\bar{z^2}}{T_d^{3/2}
{\mu^2}}}} \gamma
\rho_0^2 \left(\frac{\displaystyle \rstar}{\displaystyle \rsun}\right) \\
\end{array}
\right\}
\end{equation}
where $\lambda$ is the wavelength in cm, $g_{ff}$ and $g_{fb}$ are the
Gaunt factors for free-free and free-bound emission respectively,
$\bar{z^2}$ is the mean squared atomic charge, $\gamma$ is the ratio
of the number of electrons to ions, $\mu$ is the mean atomic mass, and
$c_{s,0}$ is the sound speed at the star.

The intensity $I_\nu(R)$ is calculated in the same way as in Waters
(1986), and hence the spectrum of the disc may be calculated. However,
here it is the total emission at a given radius which provides a
direct comparison with outflowing viscous disc models. 
The differential energy liberated is
\begin{equation}
\frac{dL_\nu}{dR} = 8\pi^2 R I_\nu(R)
\end{equation}
(equivalent to Waters' eq.12).
Finally to obtain an expression to relate to eq.\ref{dldr}, this is
integrated over all frequencies:
\begin{equation}
\frac{dL}{dR} = 8\pi^2 R \int_{0}^{\infty} I_\nu(R) d\nu.
\label{dldr2}
\end{equation}

\subsection{How important is the viscous luminosity?}

There are now two expressions for differential disc luminosity
deriving from the viscous stresses 
(eq.\ref{dldr}) and the free-free, free-bound emission of the plasma
(eq.\ref{dldr2}). Here the fraction of viscous 
luminosity to the free-free, free-bound emission of the disc is
considered. This is an 
important check on the model -- if the viscous luminosity
(eq.\ref{dldr}) is larger then the disc emission (eq.\ref{dldr2}) then
the disc cannot be driven by viscosity.
Eq.8 provides a prescription for the disc emission which may
be directly related to the observations by integrating over radius to
produce a spectrum, whereas eq.\ref{dldr}
provides the theoretical emission due to viscous dissipation.

The density parameter $\rho_0$ and the power law exponent $n$, 
can be constrained by fitting the spectrum of the disc with the
spectrum produced by Waters' method.
The ratio ${\cal F}$ of luminosity emitted from free-free and
free-bound transitions to the viscously dissipated energy is
\begin{equation}
{\cal F} = \frac{16\pi}{9 c_s^3 \rho_0 \alpha}
\left( \frac{R}{\rstar}\right)^n 
\int_{0}^{\infty} I_\nu(R) d\nu.
\label{f}
\end{equation}
The most poorly constrained parameter is then the viscosity parameter
$\alpha$ which is {\it a priori} limited to the region $0 < \alpha <
1$. However this may an underestimate if the turbulence becomes
supersonic (see \S2.4).
For the current calculation $\alpha$ is set to unity, although the limit that 
$\alpha \ltappeq 100$ is kept in mind. 


\section{A case study -- $\chi$-Oph}
To illustrate the discussion above, a test case is considered -- this
provides us with a ``real'' example.
As an example, the observations of the B2IV star $\chi$-Oph are used (Waters
1986). The star was assumed to have a mass, radius and temperature of
10.0\msun,  5.7\rstar, and 22,500K. 

\subsection{IR continuum excess fits}
\begin{figure}
\begin{picture}(100,270)
\put(0,0){\includegraphics{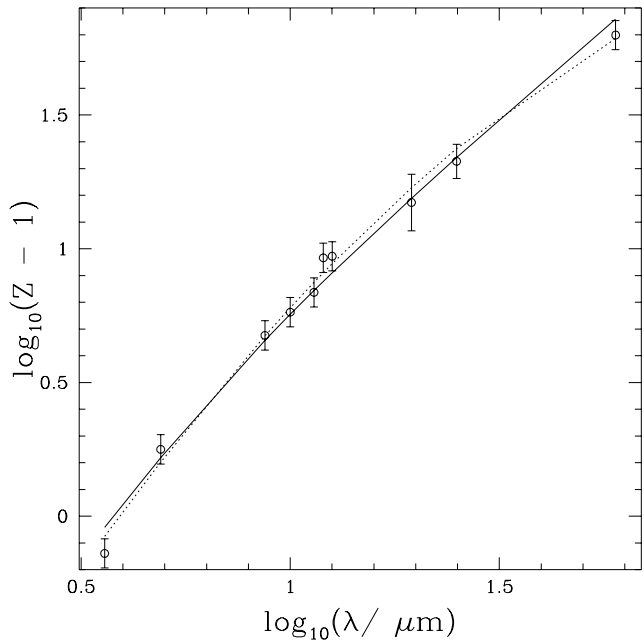}}
\end{picture}
\caption{The IR continuum excess emission as a function of
wavelength. $Z-1$ is the fractional emission excess over the stellar
photosphere. All data is taken from Waters (1986). The solid line is
the fit from the isothermal Model 1, and the dotted line corresponds to the
non-isothermal Model 2.}
\label{fig1}
\end{figure}

To constrain the power law exponents, the IR excess is first
fitted by the outflowing viscous disc outlined in \S2 using the procedure due
to Waters (1986).
Two fits to the excess were calculated: one assumed an isothermal disc
(Model 1) and the other allowed the temperature power law exponent $m$
to vary (Model 2). The disc temperature at the inner edge was assumed
to be 18000K (i.e. 0.8 of the effective temperature).  
In this latter case, the emission was calculated only for radii which
had temperatures larger than $10^4$K to ensure that the region was ionized.
In both cases the disc is assumed to have a large radius ($R_{\rm
disc} > 50\rstar$).
The best fit parameters for the models are presented in table 1, and
are shown in fig.\ref{fig1}.

\begin{table}[t]
\caption{Best fit parameters for the IR excess of $\chi$-Oph.
Model 1 is isothermal and so $m = 0$.}
\begin{tabular}{ccccc}
 & log$(\rho_0)$ (g cm$^{-3}$)  & $n$  & $m$ & $2n + 3m$ \\
Model 1 & -11.22 & 2.20 & (0) & 4.40 \\
Model 2 & -11.24 & 1.90 & 0.23 & 4.49 \\
\label{tab1}
\end{tabular}
\end{table}
Note that the derived parameters are not the same as those of Waters
(1986) as the disc model used in that work differs from that used here.
Model 2 is formally a better fit to the data -- the
reduced $\chi^2$ values are 1.1 and 0.7 for Models 1 and 2 respectively.
Although Model 2 fits the data better, it's reduced $\chi^2$ value of
0.7 indicates that it is possibly ``over fit'' and so it does not provide
strong evidence for non-isothermality in the disc. 
The temperature falls
below 10$^4$K at $R \approx 13\rstar$ providing a natural outer
boundary to the emitting part of the disc -- this radius is low when
compared to fits 
including millimeter observations of other stars (Waters \etal\ 1991).
Millimeter--centimeter data provide constraints on the extent of the
outer parts of 
the disc, and hence the temperature exponent $m$.
Column 5 of table \ref{tab1} shows the combination of $2n + 3m$ which
must be larger than 7 for a non-radiation driven outflowing viscous disc --
both models fall significantly short.

\subsection{Viscous luminosity versus free-free \& free-bound emission}
\begin{figure}
\begin{picture}(100,270)
\put(0,0){\includegraphics{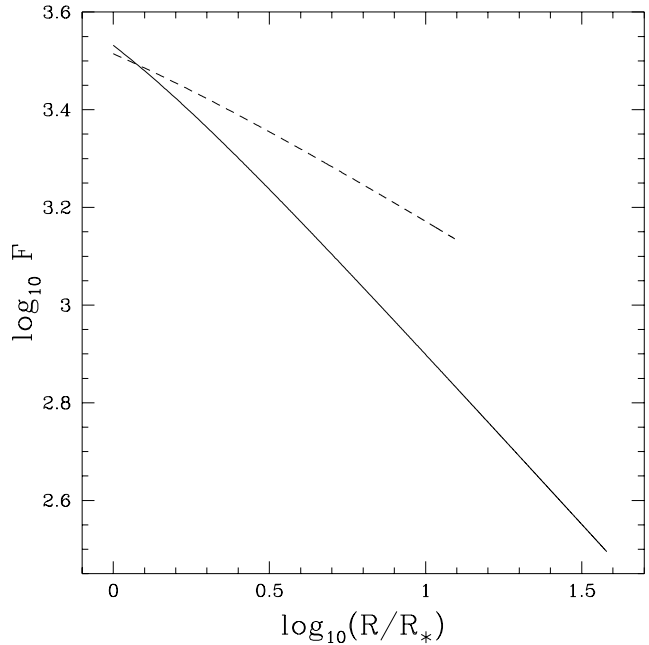}}
\end{picture}
\caption{Ratio of free-free and free-bound disc emission to the
viscous emission from eq.\ref{f}. The solid line corresponds to Model
1, the dotted line to Model 2.}
\label{fig2}
\end{figure}

Now the power law exponents have been fitted from the spectrum, the
relative contribution to the total emission of the viscously
dissipated energy can be calculated. The ratio ${\cal F}$ is evaluated
from eq.\ref{f} and is shown in fig.\ref{fig2} for both models.

Clearly the free-free and free-bound emission dominates over the
viscously dissipated energy, by of order 1000 for both models. This
result states that the energy balance within the disc is not dominated
by the energy produced by the shear motions within the disc itself,
and hence must be dominated by the stellar radiation field. 
Even with the maximum value of $\alpha \sim 100$, the free-free and
free-bound emission still dominates, making the result secure.

\section{Discussion}

The previous sections have provided a consistent description of Be
star discs as outflowing viscous discs. This study links outflowing
viscous disc 
theory with observations and finds that the two are indeed compatible
(the first outflowing viscous disc paper by Lee \etal\ 1991 produces
very high disc densities due to their choice of high disc mass-loss rate
which can be ruled out by examining the IR continuum emission).
This study therefore adds to the growing amount of evidence that Be
star discs are in reality viscously driven. There is already
substantial work regarding instabilities in Keplerian discs which give
rise to density perturbations which produce asymmetric line profiles
(e.g. Okazaki 1991). Viscous discs can now explain almost all of the
observations of Be star discs.

One major observational aspect of Be stars is that they change phase 
from Be--Be-shell--normal B star, occurring apparently at random
(although the disc may disperse and reappear in a more orderly fashion
e.g. $\mu$Cen, Hanuschik \etal\ 1993). How may the outflowing viscous
disc model explain this?
There may be a clue to these changes in the suggestions made in \S2.3
regarding the regimes where the observed discs can be outflowing discs.

Two possibilities have been speculated upon: (i)
if the disc is supported by radiation,
then it should be noted that the electron scattering optical depth is
typically in the range 0.1--1.0 (eq.\ref{tau}). 
With this so close to unity, variations in the density at the star-disc
boundary of factors of a few may be enough to decrease the radiative
driving through the disc. This loss of radiative support will lead
naturally to an accretion phase where the disc falls back on to the star. 
The density variations at the star-disc boundary could, in principle 
be created as a by product of disc warping (Porter 1998b), or
non-radial pulsations of the star (e.g. Lee \& Saio 1990). 
Alternatively, (ii) the disc may be a outflowing viscous disc due to a
possible large radial temperature gradient. However the recent
modelling of Millar \& Marlborough appears to rule this out.

With the outflowing viscous disc model able to explain the current
observations it is pertinent to ask whether future observations could
confirm the paradigm. The current major weakness with the theory
regards the input of angular momentum at the inner boundary. Although
this process has been suggested to be dissipation of non-radial
pulsations in the atmosphere (Osaki 1986), it has not been clearly
demonstrated. If this is the underlying mechanism, then there should
be statistical correlations of disc and pulsation parameters -- an
observational aspect which is still to be resolved convincingly.

\section{Conclusion}

This paper has produced several points: first, outflowing viscous
discs can only 
exist around Be stars if the disc is partly driven outwards by the
stellar radiation field
This has been discussed and shown to be
viable for actual Be star discs.

Secondly, it is found (confirming {\it a priori} expectations) that a
viscous disc can account for the observed excess IR emission of Be stars.
Also, it has been found that the energy balance in the disc is
dominated by the stellar radiation field.
The energy liberated in viscous
dissipation has been calculated and is shown to be a small fraction of
the observed luminosity of the disc ($\sim 10^{-3}$--$10^{-1}$).
These second two points are the first time it had been demonstrated
that outflowing viscous discs are energetically allowable models.

The underlying reason why Be stars go through phases where their discs
are lost has been speculated upon. This may be
due to blocking of radiative support
and driving of the disc at the inner regions -- 
causing the disc to change to an accretion disc.

\section*{Acknowledgements}
JMP is supported by a PPARC postdoctoral research assistantship.

{}

\end{document}